\documentclass[letterpaper,twocolumn,10pt]{article}
\usepackage{usenix,epsfig,endnotes}
\usepackage{amsmath}
\usepackage{verbatim}
\usepackage{xcolor}
\usepackage{subfigure}
\providecommand{\url}[1]{\texttt{#1}}

\begin{document}

\date{}

\title{\Large \bf A distributed Integrity Catalog for digital repositories}

\author{
{\rm Nikos Chondros}\\
University of Athens
\and
{\rm Mema Roussopoulos}\\
University of Athens
} 
\maketitle

\makeatletter{}\subsection*{Abstract}

Digital repositories, either digital preservation systems or archival systems, 
periodically check the integrity of stored objects to assure users of their 
correctness. To do so, prior solutions calculate integrity metadata and require 
the repository to store it alongside the actual data objects. This integrity 
metadata is essential for regularly verifying the correctness of the stored data objects.  
To safeguard and detect damage to this metadata, prior solutions rely on widely 
visible media, that is unaffiliated third parties, to store and provide back 
digests of the metadata to verify it is intact. 
However, they do not address recovery of the integrity metadata in case of
damage or attack by an adversary. 
In essence, they do not \emph{preserve} this metadata.

We introduce IntegrityCatalog, a system that collects all integrity related 
metadata in a single component, and treats them as first class objects, managing 
both their integrity and their preservation. We introduce a treap-based persistent 
authenticated dictionary managing arbitrary length key/value pairs, which we use 
to store all integrity metadata, accessible 
simply by object name. Additionally, IntegrityCatalog is a distributed system 
that includes a network protocol that manages both corruption detection and 
preservation of this metadata, using administrator-selected network peers with 
two possible roles. \emph{Verifiers} store and offer attestations on digests and 
have minimal storage requirements, while \emph{preservers} efficiently 
synchronize a complete copy of the catalog to assist in recovery in case of  
a detected catalog compromise on the local system. We describe our prototype 
implementation of IntegrityCatalog, measure its performance empirically, and 
demonstrate its effectiveness in real-world situations, with worst measured throughput of 
approximately 1K insertions per second, and 2K verified search operations per second.

\makeatletter{}\section{Introduction}
\label{sect:intro}

Digital repositories, either digital preservation systems or simply archival 
systems, store a series of objects for the long term. The former, in particular,
are expected to preserve their contents for centuries. A user of such a 
system can request a stored object a long time after its
ingestion into the archive, and expect to obtain an intact copy of the object. 

There are two key challenges a digital repository has to tackle to provide  
assurances to its users regarding the integrity of its stored objects.
The first is the physical degradation of the storage media, disk drives 
typically, that can result in bit rot and silent corruption.
The second is an adversary, external or internal, that wishes to silently alter 
the content of the stored objects, in an attempt, for example, to rewrite 
history. 

These challenges have been studied and solutions have been proposed, 
e.g.~\cite{ETSI/TS/101/903,rfc4998,HPL-2006-54,lockss-sosp-2003,conf/dgo/SongJ07}. Many
of these solutions 
generate integrity information for each object (such as a digest produced by a 
hash function), collect it in a separate file (integrity metadata file), 
and require the digital repository to store this extra file 
alongside the actual object.   The system then verifies the integrity of the data
objects stored by calculating fresh hashes of the objects and comparing them with 
the associated metadata file.  To ensure the integrity metadata is itself tamper-evident,
these systems typically aggregate a series of such metadata files into an authenticated data
structure and publish a digest of this structure in a widely visible medium, such
as a newspaper or a Usenet newsgroup.

This approach is insufficient for two reasons. 
First, introducing the use of integrity metadata to verify and preserve data objects creates
the recursive issue that integrity metadata must \emph{itself be preserved} too. Thus, it is not enough to simply make it tamper-evident; there must also be a mechanism to \emph{recover} damaged, compromised, or deleted integrity metadata.
A trivial example of an attack to silently alter content is the following: an attacker with  access to the file system deletes a series (or all) of the integrity metadata files, and then proceeds to change the few files he is interested in, rendering the repository unable to detect the exact damage caused.

Second, offloading integrity digests to third parties (e.g., Usenet)
is not future-proof.  In the long run, the organization storing the integrity
metadata summaries may go out of business, become unavailable, or simply stop cooperating.
For example, in the case of Usenet, there is no
long-term guarantee that the company storing and maintaining the Usenet newsgroup now will
continue to store integrity token summaries intact for the long term.
To truly safeguard this metadata summary, one must \emph{distribute} this information
to multiple nodes, ideally in separate administrative domains, and develop a 
protocol that manages the distribution of integrity summaries.  

In this paper, we describe the design, implementation, and evaluation of 
IntegrityCatalog, a system that can be integrated into any   
digital repository.
IntegrityCatalog takes a holistic approach.  It treats
integrity metadata as \emph{first-class objects} and serves as a tool that
ensures that the integrity of the metadata themselves is preserved, verified regularly,
and recovered in case of damage or adversarial attack.  

IntegrityCatalog stores and manages all integrity metadata indexed and accessible by
object name, alleviating the need for the digital repository to maintain and manage
extra information alongside its data objects. 
It uses an authenticated data structure 
for its storage needs that is persistent and tamper-evident.
Additionally, IntegrityCatalog is a
distributed system that includes a network protocol that manages both corruption 
detection and recovery of integrity metadata using a set of administrator-selected
network peers.  These peers have two possible roles.
\emph{Verifiers} store and offer attestations on integrity tokens and have minimal
storage requirements, while \emph{preservers} efficiently synchronize a complete
copy of the Catalog to assist in recovery in case of a catalog compromise on the local
system.
This approach allows careful selection of the entities the repository administrator entrusts to serve
as verifiers and preservers.  The repository administrator has the flexibility to
choose as many remote network peers as possible to serve as verifiers and preservers.
This will minimize the risk of a powerful adversary compromising the majority of peers, and forcing the use
of corrupt integrity information.  The administrator can also change the selection of
network peers over time as they become unavailable.

To implement IntegrityCatalog, we create a new, disk-based, tree-backed, authenticated 
data structure, called \emph{TreapPAD}. This data structure is \emph{persistent}, as 
it allows snapshots (of integrity metadata in our case) to be taken at arbitrary intervals and ensures that values 
in previous snapshots remain intact for the lifetime of the tree. It is optimized for small size, and also allows customization to
further reduce its size, by skipping authenticator caching at different depths for different snapshots, recalculating them on the spot as needed.

Our approach is simple, efficient and robust. It is simple because in contrast to existing solutions,
it aggregates all integrity information into a single (but persistent) catalog.  
It is efficient, because IntegrityCatalog verifies the complete 
data structure once, by going out on the network to verify its root authenticator, and 
then provides unlimited searches of digests within the Catalog for verification of stored data objects without requiring additional network 
communication.  Finally, it is robust because the catalog is distributed by design to avoid single points of trust and failure, and is never used without prior verification.

In summary, the contributions of this paper are:
\begin{itemize}
\item We describe the design and implementation of IntegrityCatalog, a system that can
be used as a tool by digital repositories to protect, preserve, and recover integrity
tokens in case of damage or adversarial attack.

\item We introduce TreapPAD, a disk-based persistent authenticated dictionary, 
optimized for size, with the ability to shrink it further by skipping caching of authenticators.

\item We describe the design and implementation of a protocol for efficiently distributing
IntegrityCatalog snapshots to enable preservation and verification of integrity tokens 
at remote nodes.

\item We empirically measure the performance of our IntegrityCatalog prototype using real-world data and
show that insertions into the catalog and searches/retrievals of object digests from
the catalog are efficient. 
Our experiments show worst measured performance of approximately 1K insertions per second, and search throughput of 2K verified queries per second,
when using an SSD device to store the catalog.
\end{itemize}

\makeatletter{}\section{Background}
\label{sect:backg}

Cryptographic hash functions~{\cite{STOC89*33,MenOorVan96}} map a message of  arbitrary length to a fixed size 
digest, with properties such as pre-image and collision resistance. Authenticated data structures use cryptographic hash functions to represent a 
data collection with a fixed-size authenticator. A proof for a membership 
claim, along with an authenticator, can be used by an external entity to 
verify the claim.

Merkle introduced the hash tree (or Merkle-tree) in~\cite{merkle87}, which is thought 
to be the first authenticated data structure. Here, the digest of each data 
element is hashed to produce a leaf node of a balanced tree with a fixed 
fan-out. The inner nodes of the tree are the result of hashing the digests 
produced at the lower level. With this organization, the root digest identifies 
the entire collection of data items uniquely (it is the authenticator), with the 
assumption that the above properties hold for the  cryptographic function 
used to produce the digests. The proof for a membership claim consists of all 
digests needed to re-produce the root digest, in the path from the data 
element in question to the root.

Persistence in data structures was studied by Driscoll et al. in~\cite{STOC86*109}. 
The idea is to have a linked structure (basically a tree) where changes never overwrite the older versions and one can access any previous version's values at will.

A Persistent Authenticated Dictionary (PAD)~\cite{Anagnostopoulos:2001:PAD} combines authenticated dictionaries (search trees with associated values)  with persistence.
Our TreapPAD data structure is a disk-based implementation of a PAD.

\makeatletter{}\section{System description}
\subsection{Requirements}
\label{sect:reqs}

Integrity metadata is an essential component of many digital repositories as they
perform verification of stored data objects by periodically calculating the hashes
of these objects and comparing them with the associated integrity tokens.
We treat integrity metadata as first class objects and propose IntegrityCatalog
as a tool that ensures
that the integrity of the metadata themselves is verified regularly, preserved,
and recovered in case of damage or adversarial attack.

The design of IntegrityCatalog system must fulfill the following requirements:

\begin{itemize}

\item the system should provide a succinct representation of all data objects in
the digital repository  and assist the repository in regularly verifying the integrity
of the stored data objects

\item the system should provide the repository with the modification history of
any object

\item the system should not require the digital repository to maintain any additional
information, and thus it should allow access to integrity metadata by object identifier

\item the system should manage and store all integrity metadata in a persistent,
authenticated, and tamper-evident manner

\item the system should be efficient;  insertion into and retrieval of metadata from 
the Catalog should be fast, as digital repositories often exhibit high ingress rates

\item the Catalog should not be stored solely
on the repository, as an attacker could modify data objects and integrity metadata
to match, and thus alter the content of data objects without being detected; 
the system should be able to recover the catalog to a prior consistent state in case of damage

\item to avoid dependence on a single point of trust or administration,
digests of the Catalog should be distributed to multiple nodes regularly and in
an efficient manner

\item the system should not require storage of long-term secrets, making it suitable
for use in the digital preservation domain

\end{itemize}

\subsection{Threat model}
Our primary goal is to protect the persistent state of a digital repository. 
As such, we assume an adversary that can gain access to a local node 
of the digital repository and alter its on disk data files. We also assume 
the adversary is able to do the same at some percentage of the remote peers (verifiers, preservers), who
attest to the integrity of the local catalog. 
We present thresholds for the number of remote peers that can be controlled by
the adversary in section \ref{sect:system/assurances}.
We do assume however, the attacker cannot block network communication between honest nodes indefinitely.

We assume the attacker cannot replace running or stored code silently; using Java and its code signing, for example, would prohibit this.
We also rely on the host system (the digital repository), to provide for authenticated communication channels, using some kind of PKI.
Because of this, we assume the adversary cannot communicate over said channels freely; using a pass-phrase for the local private key and providing it only when starting verified applications would prohibit this.

\subsection{IntegrityCatalog overview}
\label{sect:system/overview}
IntegrityCatalog enables a node to manage 
the association of a locally stored object (with a unique identifier) with the 
digest of its contents. The collection of associations between locally stored 
objects and their digests forms a persistent authenticated dictionary that can be represented 
by its root authenticator. We use \emph{TreapPAD}, a data structure we introduce, to manage this collection.

The contents of this collection are protected by taking snapshots at frequent 
intervals and publishing a succinct token, representing the entire collection of 
locally stored objects, to administrator-selected peers on the network. The form of this token 
is \emph{\textless snapshot\_id, snapshot\_authenticator\textgreater}, where \emph{snapshot\_id} is an ever increasing 
number identifying the current snapshot, and \emph{snapshot\_authenticator} is its corresponding 
authenticator. Some of these peers, pre-selected by the administrator, are 
expected to further contact the local node and preserve the full contents of 
IntegrityCatalog for future recovery in case of damage or attack. The system is 
optimized to minimize the interactions for preservation to a minimum, by 
allowing only pages modified since the last preservation operation to be 
transferred.

The host system using IntegrityCatalog is expected to verify the catalog once, 
before using it to access object digests. 
IntegrityCatalog contacts its peers and verifies the 
authenticator of the latest snapshot, thus certifying the entire collection is 
intact. From this point on, IntegrityCatalog provides verified replies without any further external network communication.

\begin{figure}[h]
\centering
  \includegraphics[width=0.47\textwidth]{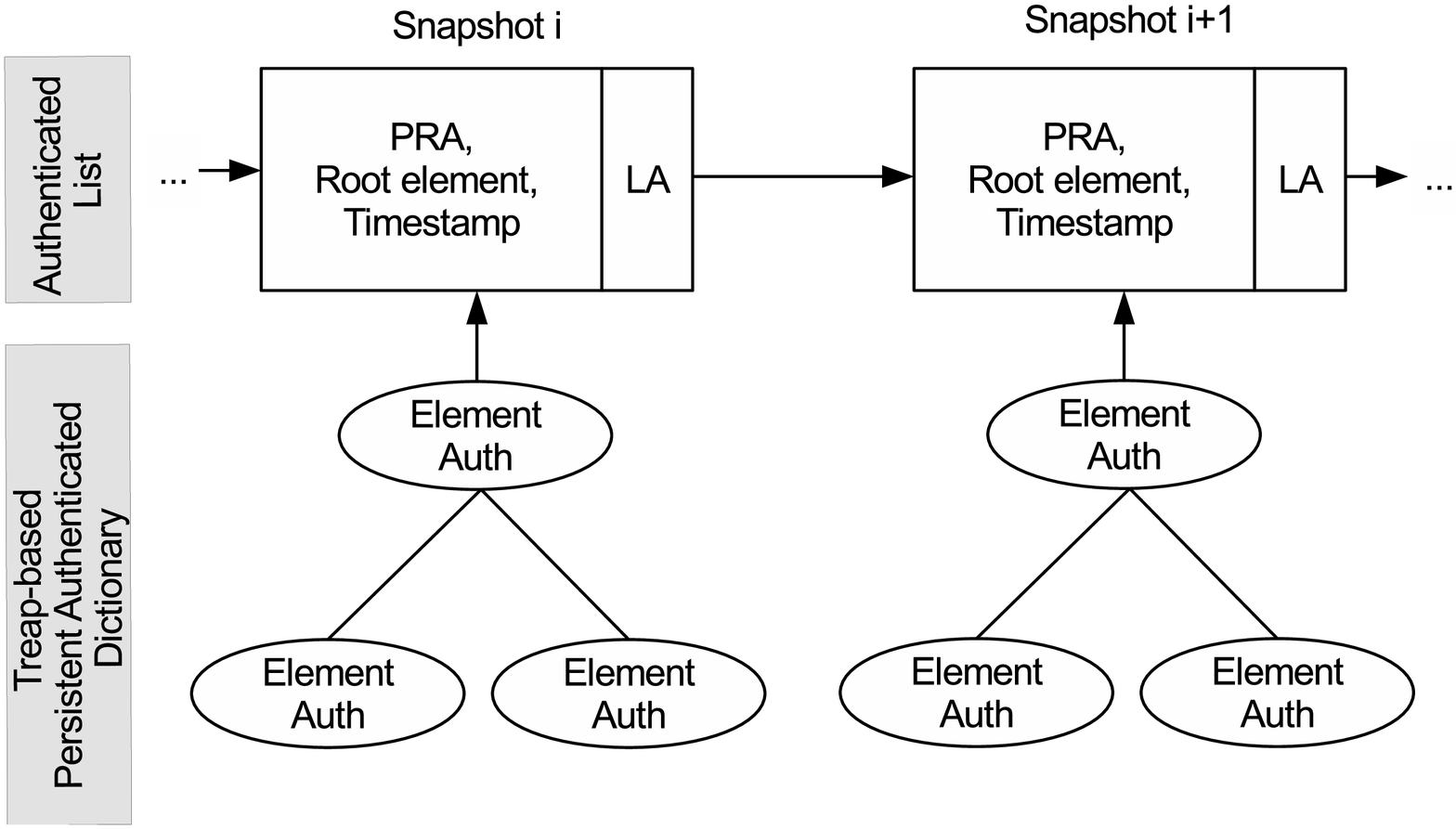}
\caption{The layout of information in TreapPAD.}
\label{figure:TreapPAD}
\end{figure}

Figure~\ref{figure:TreapPAD} shows the layout of information in the running system. 
Elements are inserted in the tree at the bottom. When a snapshot is taken, TreapPAD's Root element Authenticator is calculated (PRA) and pushed to the authenticated list, along with a pointer to the tree's current root element for this snapshot, and the system time. This produces a new list authenticator (LA) that becomes the snapshot authenticator, mentioned above, to be published to remote peers.

\subsection{Design details}
\label{sect:system/details}
\subsubsection{Integration with the host repository system}
IntegrityCatalog requires two lists of peers to be defined for its operation. 
The first one is the \emph{verifiers} list, containing nodes that are willing to 
store and later provide back the authenticator of each snapshot of the catalog. 
This information is small (a few dozens of bytes for each snapshot) imposing 
minimal storage requirements to nodes with the role of verifier.

The second one is the \emph{preservers} list, which should be 
a subset of the \emph{verifiers}, and contains nodes willing to store and potentially 
provide back the full contents of the catalog itself. As this is an ever 
growing data structure, its storage requirements are more significant than those 
ones imposed on the \emph{verifiers} for the storage of the snapshot authenticators.

The choice of nodes to act as \emph{preservers} and \emph{verifiers} is left
to the administrator of the digital repository.  Administrators have the flexibility
to choose network nodes they trust, and to change the set of selected nodes over time, 
in accordance with the repository's security policy.

The host system (i.e., the digital repository) can ask the catalog to store the digest of an object. This 
digest is expected to be calculated and formatted by the host system itself. The information 
given to IntegrityCatalog is an identifier to serve as a key and a free format 
sequence of characters (containing the digest) to serve as a value. The identifier is again treated as 
an arbitrary string, which may contain an object name, an object version or 
anything else appropriate for the underlying system.

The host system can terminate the current epoch of the catalog at intervals it 
defines itself (e.g. every day). This is achieved by invoking the \emph{sealCatalog} 
operation. This operation takes a snapshot of the catalog and produces a new 
\verb!<snapshot_id, snapshot_authenticator>! pair. It then proceeds to publish this pair 
to the nodes in the \emph{verifiers} list and waits for their acknowledgments. Once 
enough nodes have stored this information, the system considers the \emph{sealCatalog} 
operation finished and notifies the host application about the result of the 
operation. In case of error, the host application is instead notified with the 
failure.

We expect the host system to periodically use IntegrityCatalog to help it check the integrity of the objects it manages.
To do so, it requests the stored integrity information and compares it with freshly recalculated hashes from the stored object data.
Before reading the catalog, it is expected to 
invoke once the \emph{verifyCatalog} operation. This operation requests the nodes in the 
\emph{verifiers} list to transmit back the latest \verb!<snapshot_id, root_authenticatort>! 
pair each one has stored for the current node. IntegrityCatalog gathers 
all messages and tallies the replies by \verb!snapshot_id! and \verb!digest!.
It makes sure a minimum quorum of \emph{verifiers} have replied, and tests that a winning threshold of \emph{verifiers} have sent replies matching the local latest known snapshot's digest (see sections \ref{sect:system/assurances} and \ref{sect:impl/catalog} for details).

If the result of verification is successful, the host system can repeatedly 
query the catalog for the digest of any stored object to verify its 
integrity, without generating any further network communication. 
Before any reply to such queries 
is provided, IntegrityCatalog verifies it produces the same (already verified) 
root digest of the current snapshot, by rebuilding this root digest from the 
leaf containing the searched element, all the way up to the root. This way, 
any tampering of the stored data pages of the catalog itself will  be 
detected.

If, however, the verification of the catalog fails, the host system is expected 
to initiate the \emph{recoverCatalog} operation, during which IntegrityCatalog contacts all nodes 
in the \emph{preservers} list to achieve consensus on the latest available version of 
the catalog, in a manner similar to \emph{verifyCatalog} above. Once IntegrityCatalog decides on a version, it selects at random one 
of the nodes maintaining the version and requests the initiation of a binary 
update. IntegrityCatalog then restores the catalog from the selected node, one 
snapshot at a time. When finished, it notifies all nodes in the \emph{verifiers} list 
that the latest version of its catalog has been reset to the restored one. At
this point, the host system is expected to iterate over all data objects,
identify the objects missing from the restored catalog and register them. The objects
in question are the objects ingested in the local repository after the restored
snapshot of the catalog was taken. This function of the host system can be summarized as
follows:

\begin{verbatim}
for each object in repository
    hash = catalog.get(object.name)
    if hash is null
        hash = calculateHash(object.data)
        catalog.put(object.name,hash)
\end{verbatim}

\subsubsection{Change of hash function}
Every digital preservation system that uses hash functions for verifying 
integrity must address the fact that eventually the specific hash function used 
will be phased out, hopefully before it becomes compromised. 
To facilitate this, while also guaranteeing that an attacker 
cannot replace the existing digest of an object completely, 
IntegrityCatalog provides for an~\emph{amend} operation. This operation allows 
the host system to append a new value to the existing one for a specific 
object. The result of this function is that the object is associated with the 
concatenation of the old digest and the new digest supplied. It is expected that 
the host system will apply some structure to the values it puts in the catalog 
so that, over time, it can make sense of extracted information. For example, it 
can prefix the hash function name to the actual digest and append a delimiter, 
such as 'md5:XXXXXX/sha-1:YYYYYY/sha-256:ZZZZZZ'. 
In any case, the format of the value attached to each object identifier is left 
completely to the host application. IntegrityCatalog simply does not allow a 
value to be overwritten, but only amended.

In case of object migration, due for example to format change, our system 
expects the object to be re-registered, with a new version identifier. 
This version identifier is expected to be part of the object name, thus creating 
a new entry in our search tree. The idea is that all different formats exist in the catalog under different names, and most probably in different catalog snapshots.

\subsubsection{Time-stamping}
Finally, IntegrityCatalog timestamps each snapshot with the current system time. This timestamp is protected by
the same snapshot authenticator as any other data, and is used to answer modification history questions by the following approach. 
The latest change of an object is identified by the snapshot id attached to the current value as it is stored in the latest version of the catalog (called \emph{latest\_sid}). 
This may be, for example, the latest update due to hash function change. 
This snapshot id is used to obtain the timestamp of this snapshot and provide
the approximate time of latest change to this object. 
After that, this object is looked up in the catalog as it existed in the snapshot before \emph{latest\_sid}.
If it did exist, we obtain the snapshot id of its previous change, for which the timestamp can again be obtained.
This procedure is repeated until the object cannot be found before some arbitrary snapshot id, which represents its time of ingest.
The granularity of this approach is the period between snapshots, which we consider sufficient for the long-term perspective of a digital repository.

TreapPAD can easily be extended to maintain a set of values instead of a single one, for each element, making this set of values tamper-evident. These values could include further metadata, like the exact creation timestamp, file size, author, etc. We did not implement this feature as we opted for simplicity in the interface with the host system.

\subsection{Assurances}
\label{sect:system/assurances}
The basic principle behind IntegrityCatalog's security properties is never to trust information read from the local disk drives without first verifying it.
Whenever a \emph{seal} of the catalog is requested, a snapshot of the authenticated dictionary is taken and its root authenticator is calculated. 
Let's call this \emph{PAD root authenticator (PRA)}. 
This \emph{PRA} is pushed to the Authenticated List, along with the current system date and a new \emph{List Authenticator (LA)} is produced.
The \emph{LA}, along with the current \emph{snapshot\_id} form the tuple that is pushed to the remote peers.
Correspondingly, the verify operation fetches this tuple back from the remote peers.
At this point, IntegrityCatalog asks the Authenticated List for the membership proof of the retrieved snapshot id 
and verifies the extracted list authenticator against the one retrieved from the network. 
If this succeeds, it also extracts from the membership proof the \emph{PRA} for the current snapshot, without accessing the disk again, leaving no window for an adversary to alter the on-disk data structures.
It then returns these elements (\emph{snapshot\_id, LA, PRA}) to the application.

When the application tries to retrieve the security token of an object as it is stored in the catalog, it also provides the retrieved \emph{snapshot\_id} and \emph{PRA}. 
IntegrityCatalog will obtain the tree's root element for the specified snapshot, from the authenticated list, and proceed to ask the PAD for
the proof of membership for the requested element. It will verify the proof against the provided \emph{PRA} and extract from the proof the security token as it was stored from the repository, along with the id of the epoch (snapshot) it was stored. Again, no further disk access is performed with no window of vulnerability.

The same approach is used when requesting the history of changes for an object, as the snapshot id, security token, and timestamp of the snapshot are read from the membership proofs, after these are verified against the ones retrieved from the remote peers.

To summarize, IntegrityCatalog simplifies the management of integrity information in a digital repository, and provides the following assurances:
\begin{enumerate}
 \item 
If object data is corrupted, but catalog data is not, the corruption will be 
detected in the next integrity check performed by the repository.

 \item 
If catalog data is corrupted (arbitrary data pages get corrupted or purposefully 
modified but without propagating the change in the hash tree), this corruption 
will be detected on the first catalog search operation that accesses the corrupt 
data pages, raising an error. 
This detection will happen either because the page will be damaged and the 
internal tree algorithms will fail, or because the validation of the root 
digest, performed by IntegrityCatalog before returning the result, will fail.
At this point, we expect the host system to initiate a \emph{recoverCatalog} operation to 
restore an intact copy of the catalog.

 \item 
If IntegrityCatalog's data is modified in a non destructive way (from an 
attacker, maintaining the hash tree structure), its root authenticator will change 
(assumption from the definition of cryptographic hash functions). This change 
will be detected in the next \emph{verifyCatalog} operation from the peers' 
attestation. 
 
 \item 
Regarding misbehavior of remote nodes, we assume they can display arbitrary behavior and the attacker may 
control and coordinate some of these nodes.
Depending on the policy selected by the administrator of the digital 
repository, different assurances can be achieved from our network protocol.
The \emph{verifyCatalog} operation collects tuples of the form \emph{\textless 
snapshot\_id, root\_authenticator, number\_of\_votes\textgreater}, which are passed 
along with the current \emph{\textless snapshot\_id, root\_authenticator\textgreater} data 
from the catalog's tree to a policy-based checker. 

An attacker wishing to enforce an illegitimate change in an object, will first need to break into the digital repository and alter the object's contents and the catalog's corresponding entries. Additionally, he will need to control enough nodes in the \emph{verifiers} set that will: a) force a big enough quorum, to pass the quorum test, and b) force an altered catalog's authenticator to win the voting.
That is, he will need to ensure the following two inequalities hold:\\
\begin{subequations}
\begin{flalign}
 \frac{verifiers - absentees}{verifiers} > quorum\% \label{ineq1} \\
 subverted > winning\% * (verifiers - absentees)\label{ineq2}
\end{flalign}
\end{subequations}
where $absentees$ represents the number of faulty non participating nodes, $subverted$ is the number of faulty participating nodes, $quorum\%$ and $winning\%$ are the percentages selected by the current policy, and $verifiers$ is the number of verifiers.
Inequality \eqref{ineq1} holds when the attacker passes the quorum test (we assume he can prevent nodes from participating in the voting process), while \eqref{ineq2} holds when he controls the majority of votes. 
Concluding, as long as the attacker cannot enforce inequality \eqref{ineq2} above, he cannot force the use of corrupt catalog information from network peers.

\end{enumerate}

\subsection{Limitations}
\label{sect:system/limitations}
Even with all the assurances IntegrityCatalog provides, two 
windows of vulnerability remain for an attacker to modify an object's content and go 
unnoticed. The first one is on ingress, while the current catalog epoch is 
still open. Before taking the snapshot and publishing the catalog's 
authenticator to its peers, an attacker that has taken over the local node can 
alter both the object and its digest stored in IntegrityCatalog (e.g., via 
direct binary modification of the catalog's on disk image).
One solution for this is to shrink the window 
between snapshots, at the expense of increasing the amount of network traffic as 
well as the storage requirements for the nodes in the \emph{verifiers} list.

Additionally, an adversary can attack immediately after a recovery of the 
catalog. Assuming an object was initially registered in a recent snapshot,
and this snapshot has not yet propagated to the \emph{preservers}, the attacker can 
corrupt the catalog and wait for the recovery of a previous version of it, where
the target object was not yet included. The attacker can then modify the object
before the host system re-registers it in the catalog; this will result in the object being 
registered with the wrong (altered) contents. 
This way however, the attacker can only target fairly recent objects.  Otherwise, 
he would need to control several nodes in the \emph{preservers} list, to be  able to enforce a 
version of the catalog old enough to not include an older object.
 
\makeatletter{}

\section{Implementation}
\label{sect:implementation}

\subsection{TreapPAD}
\label{sect:impl/TreapPAD}

The IntegrityCatalog prototype is implemented in Java without any external library dependencies.

Initially, we build a variable length record manager, based on the approach from~\cite{conf/adc/ZobelMS93}.
The basic idea is to use fixed-length blocks to efficiently manage variable length entities, 
which in our case correspond to the nodes of our tree based PAD.
Then, we build our persistent authenticated dictionary over a deterministic treap, which is a 
randomized search tree~\cite{SeiAra96}. We replace the random \emph{priority} of the randomized tree, 
with the hash value of the key, making it deterministic. We choose this data structure over Red-Black
or AVL trees, even though the latter are better balanced, because of its set-uniqueness property; i.e., 
the fact that no matter in which order one adds the elements, the resulting tree is the same.
We need this property for our efficient network synchronization.

We build partial persistence in our treap using a bounded fat node with node copy on overflow. More specifically, 
we maintain all persistence information in the tree node until the node expands in size over the 
maximum entry size of our variable length storage manager. At that point, we copy the node and fix
the path from the root accordingly. That is, instead of bounding the node over a fixed number of elements for each
of the left pointer, right pointer, value, and authenticator placeholders, we choose to bound the complete 
structure over its total size, while it fits in our lower level storage mechanism. Our bounded fat node approach is size
optimal, as key/value pairs are infrequently copied; in contrast, in node-copy approaches to persistence, multiple copies would be generated
resulting in increased size of the data structure.

Using these primitives, we build our TreapPAD which can store all entries requested by the repository, and also
take snapshots when requested. These snapshots cause a new set of authenticators to be generated for 
each entry that was newly inserted, as well as all of its ancestors. This generates a lot of 
authenticators for each snapshot, especially for entries near the root of the tree. 
Here, a trade-off arises.
While at one extreme, every authenticator could be cached 
, these entries are actually optional and can be recalculated at will, with the corresponding
computing speed cost. At the other extreme, one can avoid caching authenticators altogether, 
at the cost of significant recalculation effort for each attempt to access these authenticators (while building membership proofs).
We introduce the authenticator cache selective storage subsystem to manage this trade-off,
by selectively caching authenticators at different depths of the tree. 
This subsystem can be initialized with a single integer skip number that works as follows.
A skip number of 0 implies storing all authenticators. Any other positive number is used along with the current snapshot
number (1-based) and depth (0-based) to calculate whether the current authenticator should be cached (stored) using the following formula:
$$depth \bmod skip\_no = (snapshot - 1) \bmod skip\_no$$
For example, if \emph{skip\_no} is 3, the first snapshot has authenticators cached at nodes of depth 0, 3, 6, ...;
the second snapshot at depths 1, 4, 7, ..., and so forth.
With this approach, a skip number of $i$ may result in recalculation of up to $2^i$ missing authenticators.
We measure and present this attribute's impact on performance in Section \ref{sect:evaluation}.

Additionally, we need an authenticated list to track the pointer to the current root node, the root authenticator, and the current timestamp for each snapshot.
We use the Authenticated Append-Only Skip List (AASL) from~\cite{SECURITY`02*297}, to do so.

Finally, we enable efficient synchronization of a tree with an existing previous version of it.
This allows us to propagate IntegrityCatalog's underlying data structures 
to selected peers on the network efficiently, as we describe in the network operations of Section \ref{sect:impl/catalog}.
The TreapPAD interface includes the following methods to allow for this synchronization:
\begin{itemize}
\item getFirstBlockOfSnapshot(long snapshot) $\rightarrow$ $<$blockNumber, blockData$>$
\item getNextBlockOfSnapshot(long snapshot, long blockNumber) $\rightarrow$ $<$blockNumber, blockData$>$
\item binaryUpdateBegin(long snapshot)
\item binaryUpdateBlock(long snapshot, long blockNumber, MemoryBuffer blockData)
\item binaryUpdateCommit(long snapshot)
\end{itemize}

To explain, assume an application that has access to two trees, one called 
\emph{updated} and the other \emph{stale}, with obvious roles. The application will 
start by asking the \emph{updated} tree for the first block of the next snapshot after 
the latest of the \emph{stale} tree (getFirstBlockOfSnapshot). It will initiate an 
update operation on the \emph{stale} tree (binaryUpdateBegin) and update it with the 
retrieved block (binaryUpdateBlock). It will then ask the \emph{updated} tree for the 
next block (getNextBlockOfsnapshot) until it receives an End-of-Data reply, at 
which point it will conclude updates to the \emph{stale} tree (binaryUpdateCommit). It 
will continue this update loop for any snapshots existing at the \emph{updated} tree 
and not yet present on the \emph{stale} one, until the last one is processed. At this 
point, the \emph{stale} tree will be semantically equal to the updated one.

\subsection{IntegrityCatalog}
\label{sect:impl/catalog}
IntegrityCatalog exposes a single class named \emph{Catalog} that provides access
to all described functionality via its exposed methods. The basic operations,  such as 
\verb!put(key, value)!, \verb!amend(key, new_value)!, and \verb!get(key)! are self explanatory. 
We now describe the seal, verify, recover, and preserve methods.

\textbf{seal:} The \verb!seal()! method terminates the current epoch of the catalog and notifies \emph{verifiers}. 
Initially, it derives a new snapshot. After that, it sends a \verb!StoreMessage! to each Node in the 
\emph{verifiers} list, carrying the new snapshot id and its root authenticator. Finally, it 
sets a timeout to limit the time window for arrival of the replies. Each 
receiving Node is expected to store this information and reply with a 
\verb!StoreReplyMessage!. Additionally, if the \emph{verifier} node is also a \emph{preserver}, it 
is expected to schedule a remote catalog preservation asynchronous operation 
(explained in detail below) to preserve the current version of the catalog. Each 
\verb!StoreReplyMessage! that arrives at the current node is collected; once all nodes 
have replied or the timeout has passed, the system counts all positive replies and 
decides on the success of the operation, based on the defined quorum percent. 
The function used to decide the result is 
replaceable by the host system and, as such, a custom policy may be implemented.
Finally, the host system is notified with the seal operation's result.

\textbf{verify:} The \verb!verify()! method verifies the stored catalog contents are intact. 
The host application is expected to call it before looping through 
its local objects and accessing the get() function repeatedly to perform integrity checking of its locally stored
data objects. Initially it 
sends a \verb!StoredVersionRequestMessage! to each Node in the \emph{verifiers} list 
and sets an appropriate timeout. Each receiving node is expected to access its 
local registry of stored \verb!<snapshot_id, root_authenticator>! tokens for the 
requesting catalog and reply with the most recent one via a 
\verb!StoredVersionReplyMessage!.
The local node collects these replies, aggregates them by \verb!<snapshot_id, root_authenticator>! and passes them to the policy based function, along with the current tree snapshot and digest. 
This function will reply with either success or failure.
Our default policy decides success when the quorum is greater than $50\%$ of the number of \emph{verifiers}, and the current 
\verb!<snapshot_id,root_authenticator>! has received a minimum of $70\%$ of the votes of this quorum.
Again, the host application is notified of the result.

\textbf{recover:} The \verb!recover()! method restores the catalog to a previous stable version. 
The host application should invoke it only when the catalog is 
considered damaged, as it will result in the complete replacement of the local 
catalog with the latest available version from one of the preserving nodes. This 
operation takes place in three consecutive asynchronous phases. In the first 
phase, a \verb!RecoverVersionRequestMessage! is sent to each Node in the 
\emph{preservers} list. Each receiving node is expected to access the actual catalog it 
preserves for the calling node, obtain the latest available snapshot id and its 
accompanying root authenticator and send it via the \verb!RecoverVersionReplyMessage!. 
The originating node will gather all replies and decide on the latest version to 
restore, in a manner similar to the one described in the \emph{verify} function. 
Our default implementation will choose the version with more than $70\%$ of the replies, with a minimum quorum size of $50\%$ of the number of \emph{preservers}. 
It will then pick at random one of the nodes maintaining the selected version 
and start the second phase, the actual catalog recovery. 
It will send a \verb!RecoverBeginMessage! to the selected node, starting with the first snapshot id, while 
also recreating the local catalog from scratch. The receiving node will retrieve the first available block for the given 
snapshot of the preserved remote tree and reply with it via a \verb!RecoverDataMessage!.
The local node will update its catalog with the received data and then continue 
with a \verb!RecoverGetNextMessage! until it receives a \verb!RecoverEndOfData! message. At 
this point, it will increase the snapshot id and restart this phase until the 
catalog arrives at the target snapshot id.
Once this goal is achieved, at phase three, it will verify the catalog's current authenticator against the expected one and send a 
\verb!VersionResetMessage! to all \emph{verifiers}, instructing them to force the 
latest version to become the one just restored, regardless of each node's local 
registry.

\textbf{preserve:} As already stated, the remote catalog preservation process is initiated when a 
remote node (called \emph{preserver}) receives a 
\verb!StoredVersionRequestMessage!. 
In this case, \emph{preserver} will open the catalog stored 
locally that corresponds to \emph{origin} and retrieve the latest snapshot id. 
It will then proceed to send an \verb!UpdateBeginMessage! to \emph{origin} 
requesting the succeeding snapshot from the discovered one. 
\emph{Origin} will process this message and reply with an 
\verb!UpdateDataMessage! with the contents of the corresponding page of its 
catalog. \emph{Preserver} will then send an \verb!UpdateGetNextMessage! until it 
receives an \verb!UpdateEndOfDataMessage! from origin. Finally, \emph{preserver} 
will repeat this process until the target snapshot id has been derived.

\subsection{Planning for the future}
\label{sect:impl/future}
While our solution is modest in its size requirements, demonstrated to be O(N) in the number of elements of the PAD in our evaluations,
its growth is still constant over time. After enough years, a single large file may prove unmanageable as insertion efficiency degrades.
One way to address this is via partitioning at the tree level; while we have not implemented it yet, we describe it here for completeness.
Basically, instead of having one TreapPAD component for the catalog, one can employ $k$ TreapPADs and a partitioning function that, given the key
of the item to insert, assigns it to one of the $k$ trees. A very simple such function, for $k=256$, simply takes the last binary byte of the hash value of the object's name (an operation needed anyway, to calculate the \emph{priority} for our treap). During insertion, the partitioning function is invoked with the key and the insert operation is routed to the appropriate tree. 
To take a snapshot of the tree, a snapshot operation is invoked on each one of the $k$ independent trees and the root digests are gathered in one flat authenticated list. This list's authenticator becomes the root authenticator for the current snapshot and can be used on all operations.
The main gain from this approach is the reduction of the $k$ different trees' depth by a factor of $log_2 k$, resulting in fewer disk accesses.
An extra gain from this partitioning, is the ability to have up to $k$ parallel operations (insert or search) running, without introducing synchronization in TreapPAD's code, by simply maintaining $k$ different and parallel operation queues. 
Of course, small adjustments will be needed in the \emph{preserve} and \emph{recover} network operations as well.

\makeatletter{}\section{Evaluation}
\label{sect:evaluation}

\begin{figure*}[ht]
 \centering
  \subfigure[Insertion]{\label{figure:HDD_Insertion}\includegraphics[width=0.33\textwidth]{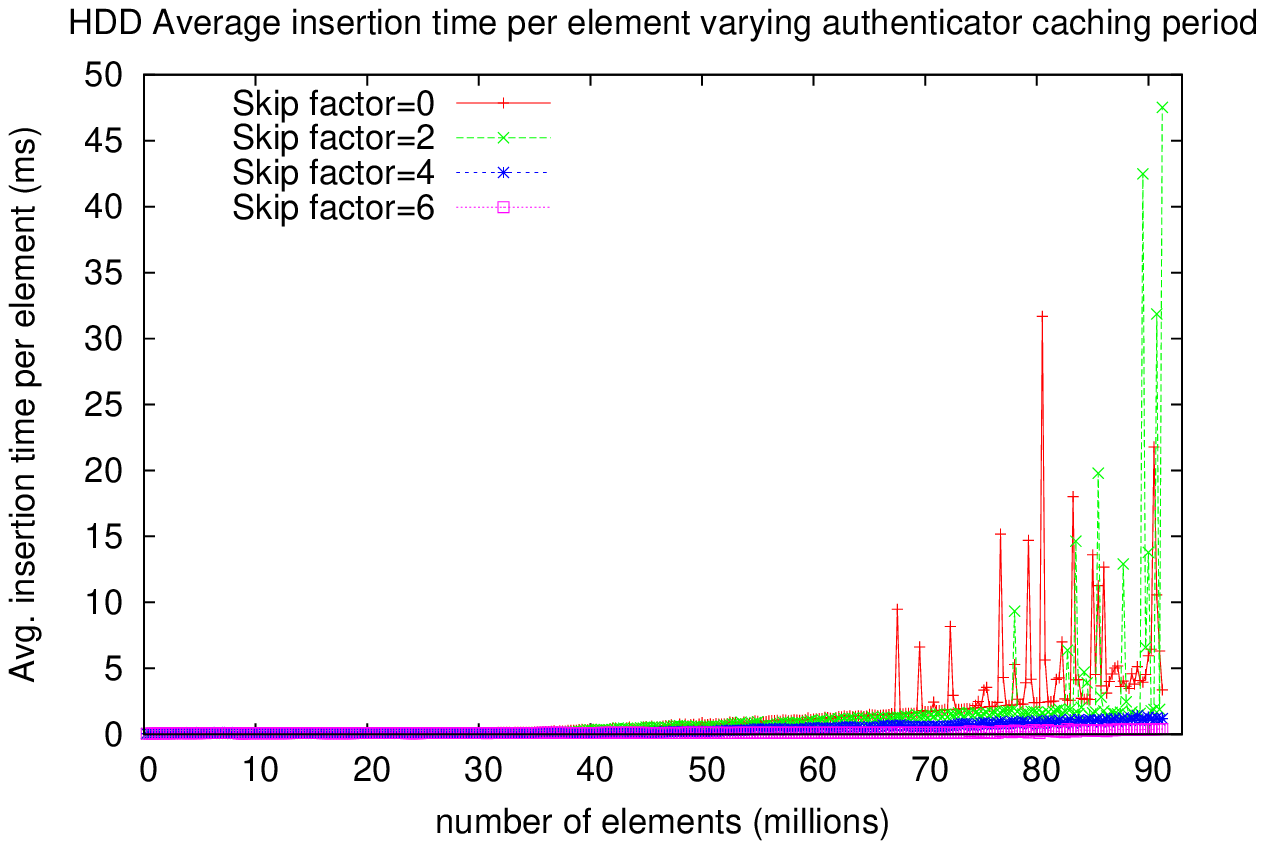}}
  \subfigure[Snapshot]{\label{figure:HDD_Snapshot}\includegraphics[width=0.33\textwidth]{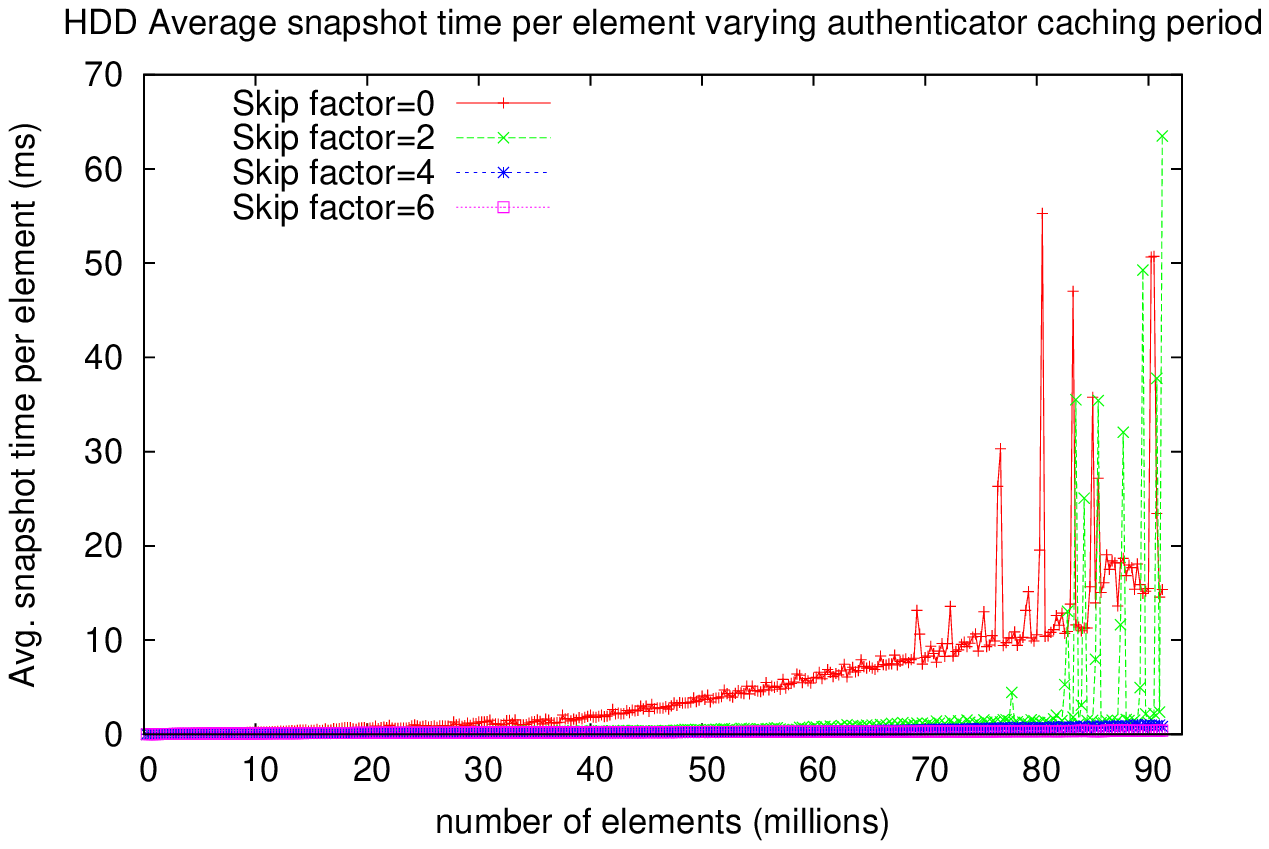}}
  \subfigure[Search]{\label{figure:HDD_Search}\includegraphics[width=0.33\textwidth]{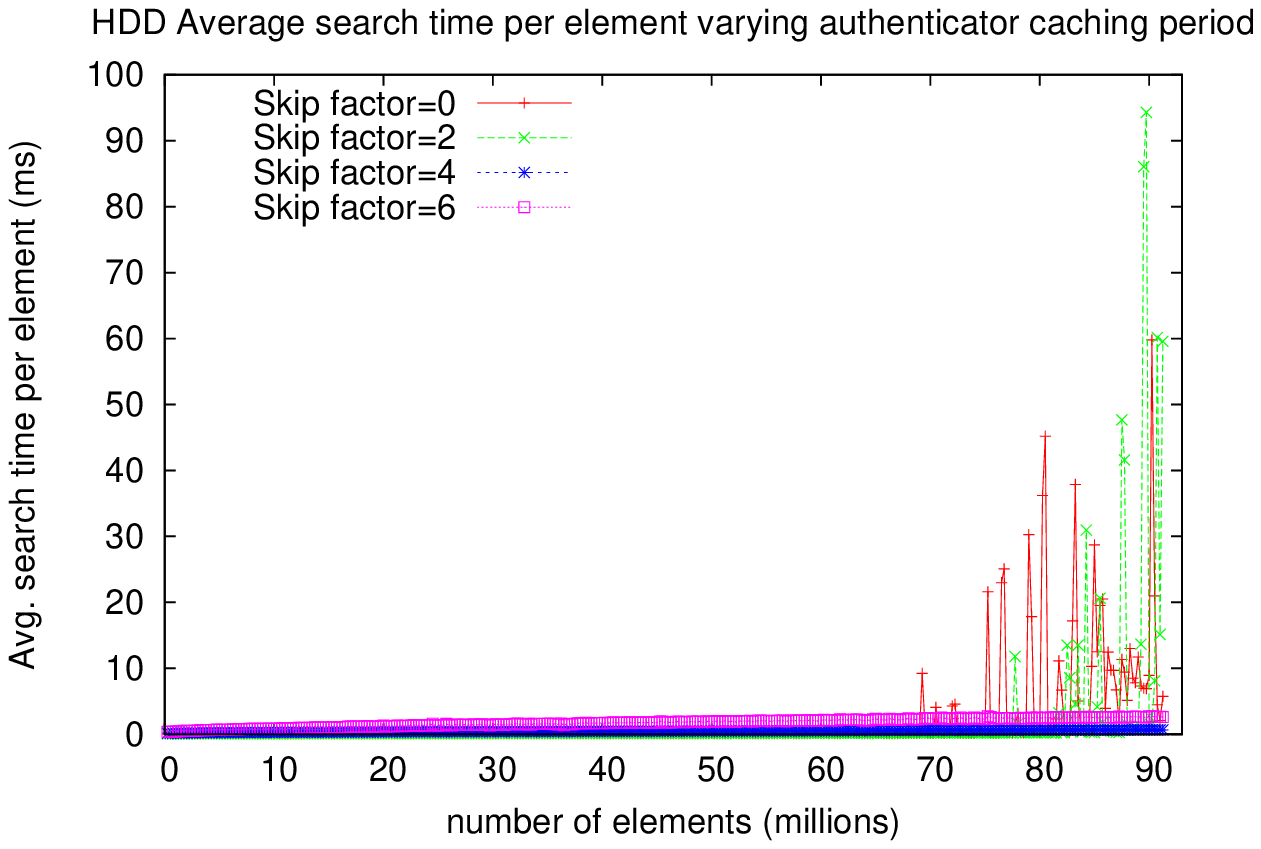}}
  \caption{Average per-element performance of insert, snapshot and search while using a HDD to store the TreapPAD, varying the authenticator caching skip factor.}
  \label{figure:KPS}
\end{figure*}

\begin{figure*}[ht]
 \centering
  \subfigure[Insertion]{\label{figure:DEVCROSS_Insertion}\includegraphics[width=0.33\textwidth]{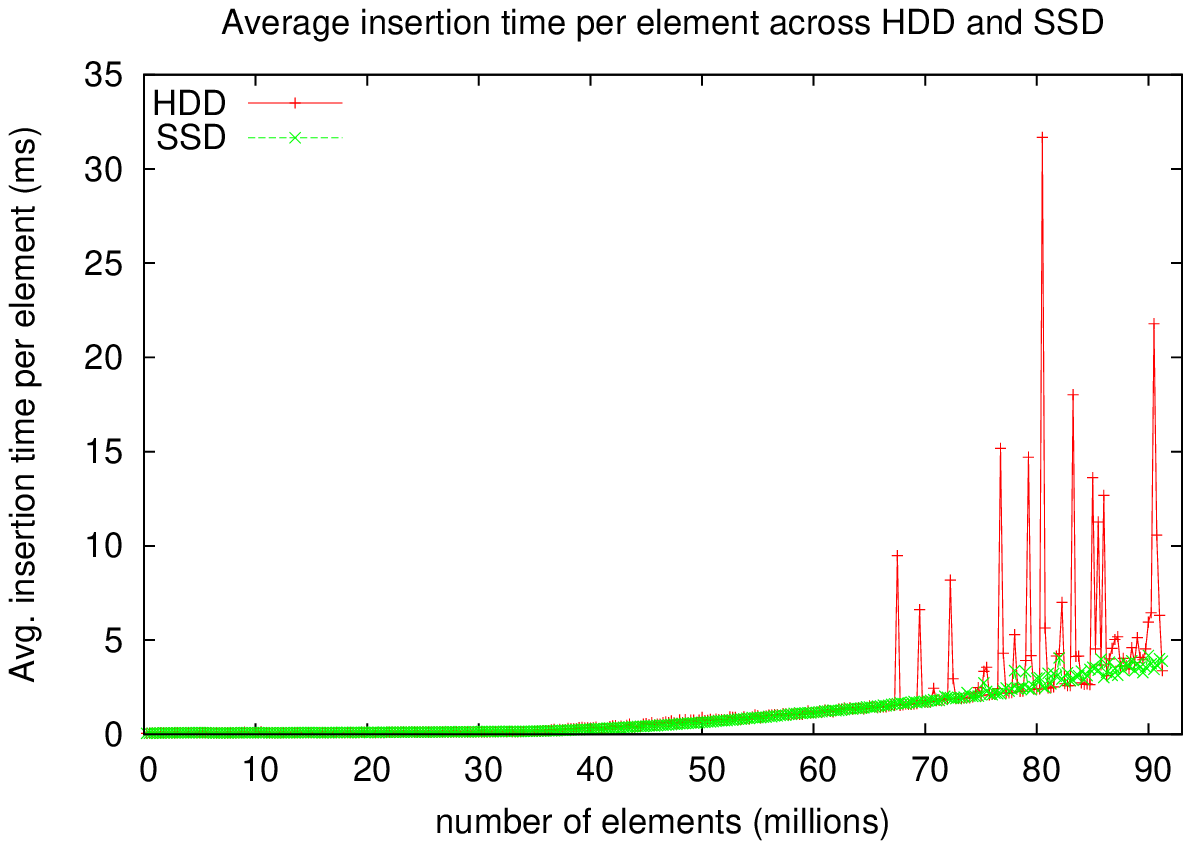}}
  \subfigure[Snapshot]{\label{figure:DEVCROSS_Snapshot}\includegraphics[width=0.33\textwidth]{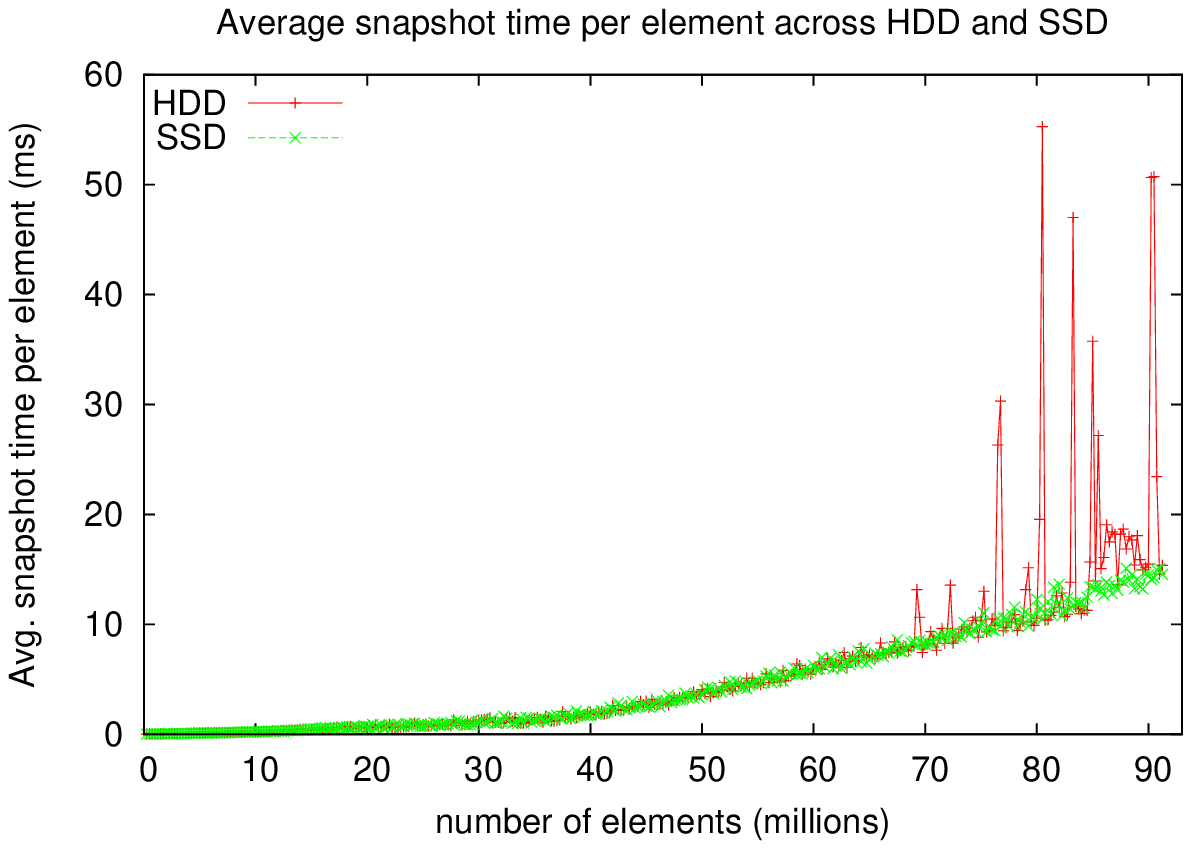}}
  \subfigure[Search]{\label{figure:DEVCROSS_Search}\includegraphics[width=0.33\textwidth]{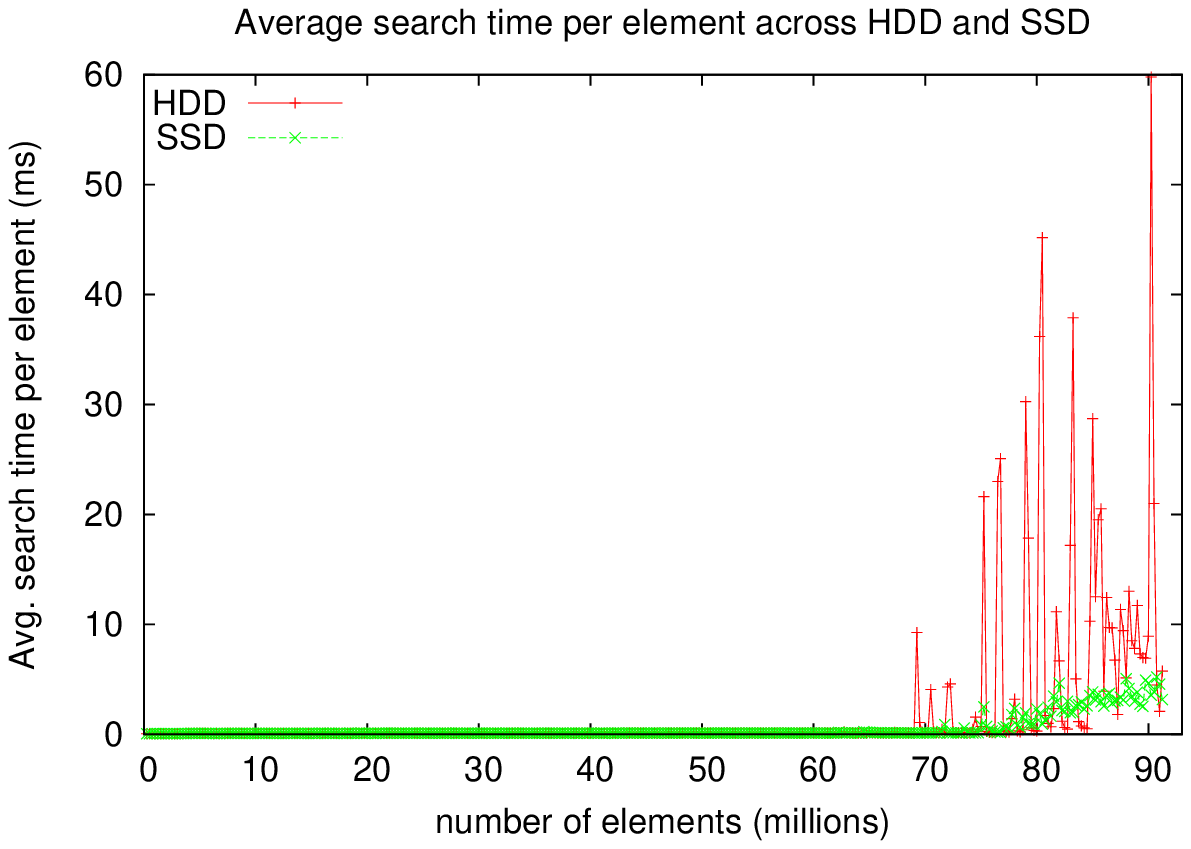}}
  \caption{Average per-element performance of insert, snapshot and search, with a skip factor of 0, HDD vs SSD.}
  \label{figure:KPS}
\end{figure*}

In this section, we evaluate IntegrityCatalog's performance. We  
focus on the performance characteristics of TreapPAD, as the 
major IntegrityCatalog operations (insert and search) are dominated by its behavior. In all 
tests, we use a single computer with an Intel Core i7-4820K running at 
3.70GHz, 64GB memory, 8 TB of hard disk data (2 x Western Digital Black Series 4TB, 7200 RPM, no RAID), and 1 TB of SSD (Samsung 840 Evo Series).

We use real-world data for our tests. Since we expect our system to be used with file names and web URLs, we chose to focus on the latter as it is more universal. To this end, we downloaded the index of the ClueWeb12 data set of the Lemur project~\cite{ClueWeb12/site} and selected
91.250.000 distinct URLs from it at random (average URL name length 51, maximum 2K bytes). 
This number represents a full year's worth of crawling (365 days x 250K URLs/day) for a single busy node 
of the live LOCKSS~\cite{lockss-sosp-2003} network (David Rosenthal, private communications), and we make the conservative assumption, that every crawled page has been modified and needs to be inserted in the archive as a new version.

Our methodology is as follows: we insert 250K URLs in TreapPAD and then we take a snapshot, simulating one snapshot per day. At this point, we search for 100K URLs at random, from the ones already inserted at any snapshot, simulating a verification of a subset of the archive. The search time includes the descent into the tree to locate the element (while also building the proof, by incorporating all necessary neighbor elements), as well as to verify the proof against a target root authenticator (recalculating all hashes up to the root). We run our tests against both a hard disk and a solid state disk. We also vary our authenticator caching skip factor from 0 (caching all authenticators) to 6 with a step size of 2. For all our tests, we use a page size of 16KB, a value we derived as optimal via a separate set of micro-benchmarks, omitted for brevity.

\begin{figure}[h]
\centering
  \includegraphics[width=0.47\textwidth]{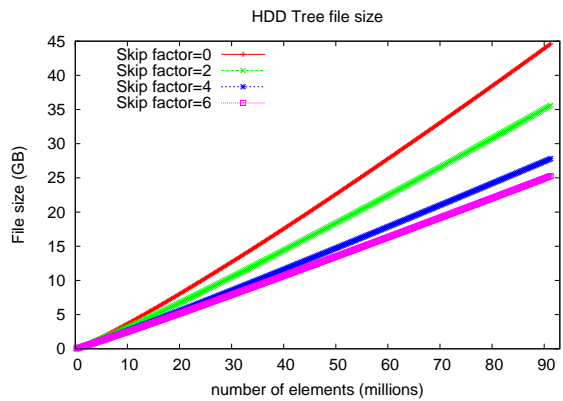}
\caption{File size measured at every snapshot point across authenticator caching skip factors}
\label{figure:size}
\end{figure}

In Figure \ref{figure:size}, we present the catalog file size as measured after each snapshot across different authenticator caching skip factors. We see size increases linearly with increasing number of elements, while it also decreases as we increase the skip factor. In the worst case, after simulating a year of operations, file size is measured at 45GB, while with a skip factor of 6, it shrinks to 25GB, almost half the size. For comparison, the text file with all the URLs used in this experiment is approximately 6GB in size, which excludes 25 bytes per URL we add representing the hash value (approximately another 2GBs).

In Figure~\ref{figure:HDD_Insertion}, we present the average time of insertions when using the hard disk, varying the skip factor. For the first 65M elements all options perform on par with sub-millisecond insertion time. At this point, for skip factors of 0 and 2, the buffer cache of the machine is exhausted and operations start fetching data from the hard disk. The other two runs with higher skip factors, due to their smaller on-disk image, still operate from the buffer cache. Even when using the disk, the worst insertion time measured is 31 ms. This corresponds to a throughput of 667 insert ops per second.

In Figure~\ref{figure:HDD_Snapshot}, we present average snapshot time (the elapsed time to create a snapshot divided by the number of new elements added). As with insertions, we see a similar pattern regarding system buffer cache usage while the file size is small enough.
The experimental run storing all authenticators (skip factor=0) exhibits a steady, sub-linear, decrease in performance with increasing file size. 
With the disk involved, the maximum average time is measured at 55 ms, corresponding to a throughput of 183 snapshotted elements per second. 
The run with a skip factor of 2 performs slower after 80M elements, while the other two still operate from the buffer cache with sub-millisecond average per element speeds.

The performance penalty of increasing the skip factor is visualized in Figure~\ref{figure:HDD_Search}, where we plot average search speed. While increasing the skip factor to 4 decreases the search time, as the complete tree file lies in main memory, with a skip factor of 6 the time actually increases, as up to $2^6$ non-cached authenticators need to be recalculated. For skip factors greater than 6, this trend will increase exponentially. Concluding, a skip factor of 4 seems to indicate a sweet spot that balances size and speed.

To understand the impact of the hard disk seek time on our prototype's performance, we repeat all experiments on the SSD. 
We plot average insertion, snapshot, and search time for HDD and SSD, for the largest file size with a skip factor of 0, in Figures~\ref{figure:DEVCROSS_Insertion} to~\ref{figure:DEVCROSS_Search} respectively.
The figures show a significant improvement in performance on the SSD, because disk seek time is eliminated. 
With the modest file sizes we presented above, we conclude it is both feasible and efficient to use an SSD for IntegrityCatalog.
For reference, we list the worst measured performance in operations per second, in Table~\ref{table:tps}, for both HDD and SSD experiments.

\begin{table}
  \begin{tabular}[h]{c r r r r r r}
     & \multicolumn{2}{c}{Insert}& \multicolumn{2}{c}{Snapshot} & \multicolumn{2}{c}{Search}\\
    SF &HDD&SSD&HDD&SSD&HDD&SSD\\
    \hline
    0 & 667 & 986 & 183 & 217 & 537 & 1889\\
    2 & 821 & 4222 & 699 & 1711 & 601 & 3321\\
    4 & 2766 & 7798 & 3204 & 4783 & 2079 & 2032\\
    6 & 6577 & 8234 & 3684 & 4184 & 534 & 534\\
  \end{tabular}
  \caption{Worst measured performance in operations per second. SF is the authenticator caching skip factor.}
  \label{table:tps}
\end{table}

Regarding our network protocol, the network traffic generated for each 
operation, per node, is presented in Table \ref{table:network} where we  
assume a SHA-1 hash function (with a digest size 20 bytes). The frequent 
\emph{seal} and \emph{verify} operations are very efficient as only digests 
travel across nodes, while the heavier \emph{preserve} and \emph{recover} are 
dominated by the size of the snapshot and tree size they carry, respectively. 
For \emph{preserve} and \emph{recover} operations, 
every network message carries a single tree page. 
Analyzing a bit further the 45 GB catalog size (the worst case above), we see it represents an average 126 MB per snapshot, the data expected to travel on the wire for a synchronization operation of an up-to-date \emph{preserver} node. At a modest network speed of 10 Mbit/sec, it will take an average of 2 minutes for a (fairly common) snapshot transfer, and a total of 13 hours for a complete catalog restore in the (rare) case of a needed recovery. 

\begin{table}
  \begin{tabular}[h]{l c c }
    operation & local & remote\\
    \hline
    seal & \textless 100 bytes& \textless 100 bytes\\
    verify & \textless 100 bytes& \textless 100 bytes\\
    preserve & snapshot size & small (ACK)\\
    recover & small (request) & catalog size\\
  \end{tabular}
  \caption{Network traffic generated by the catalog.}
  \label{table:network}
\end{table}

\makeatletter{}\section{Related work}
\label{sect:related}

One way to safeguard the integrity of an object is to use a Redundant Array of Inexpensive Disks 
(RAID)~{\cite{patterson88:raid,ChenEtAl94,bitton88,park:pario}} where either full
copies are maintained or parity bits are automatically calculated 
and stored along with data, allowing recovery as long as a minimum number of 
disks are intact. However, a significant space overhead is imposed to achieve 
integrity. Still, this approach helps repair damage besides detecting it.

Another approach is to calculate a checksum for an object and store it along 
with it. To achieve this, one can use error detecting techniques~{\cite{Hamming50}}, such
as cyclic redundancy checks~{\cite{Peterson72}}, such as the widely used 
CRC32~{\cite{conf/dsn/Koopman02}}. Although they may be attractive for messages
on communication channels and very fast to compute, they are not appropriate for long term storage,
as they do not provide strong pre-image resistance (an attacker can fairly easily calculate
a second message with the same CRC as an existing one).

Another alternative for a ``summary'' function are algebraic formulas that produce 
signatures~{\cite{Harrison71,Rabin81,KarRab87}}.
Schwarz et al.~{\cite{Schwarz06}} proposed their use for proof of remote
data possession when the owner no longer holds the original data. 
Although potentially faster to calculate
than cryptographic hash functions, they also lack the pre-image resistance properties 
required to be applicable to the long-term digital preservation and archival
systems we target.

Cryptographic hash functions~{\cite{STOC89*33}}, such as 
\emph{SHA-*}~{\cite{FIPS:2002:SHS}}, may be used to produce strong digests. This is today the strongest type of 
summary information that can be produced, and this is the scheme upon which we build.

Fault-tolerant file systems, such as Oracle's ZFS~\cite{Zlotnick06}, combine many of the above techniques for safe-guarding against hard disk drive faults.
However, they are not designed to protect against different machine-level faults, such as main memory errors, as demonstrated in~\cite{conf/fast/ZhangRAA10}.
Moreover, ZFS as well as all above approaches, where all integrity checking related artifacts are 
stored locally (either multiple copies of an object, or its checksum/digest), fail 
when a malicious entity attacks aiming to alter the existing content unnoticed. 
The attacker can simply alter all related elements and the changes
will go unnoticed by any of the above techniques.

A different approach is demonstrated by the LOCKSS~{\cite{lockss-sosp-2003}} 
peer-to-peer digital preservation system, which assumes multiple nodes store the 
same object and uses them for verification (and also for damage repair). 
This is a valid assumption for this 
system as its purpose is to preserve academic journals, that is, widely visible 
data. Each node verifies the integrity of its stored data objects by 
initiating a poll per object in which other nodes participate.
The participating nodes need to read and process the complete object 
to produce a digest for it. This results in heavy load for the system 
as a whole; our approach can complement the existing polling protocol by 
allowing each node to verify its own content. This can reduce the polling rate 
per object and thus reduce overall system load.

Haber et al.~{\cite{HPL-2006-54}} propose 
a Content Integrity Service (CIS) based on hash trees along with hash-linking. 
The system defines epochs of operation. At the end of each epoch, the root 
digest, which is dependent upon the one of the previous epoch, becomes the 
'witness' value and is expected to be safely published to a ``widely available 
medium''. 
The CIS scheme was later enhanced by Song et al.~{\cite{conf/dgo/SongJ07}} in the 
ACE system by introducing another epoch, above the aforementioned one, before 
generating the witness value. More specifically, the first epoch, which is 
dynamically sized by the system, forms a hash tree on ingest. Once it is 
closed, the root value is incorporated in a hash-chain for the user-defined period 
(e.g. a week) of the second epoch. When this second epoch closes, the 
authenticator for this hash chain becomes the witness value to be published, or 
in any case preserved as securely as possible.

The major drawback of the CIS and ACE systems is that the repository is asked to 
safely preserve the membership proofs for each data element while the integrity 
service only maintains hash trees of digests
. Thus, the simplest form of attack for an adversary is to erase 
the membership proof of an object, and then proceed to modify it, leaving the 
repository unable to verify the object's integrity.

To address this, the European Telecommunications Standards Institute (ETSI) defines
a specification~{\cite{ETSI/TS/101/903}} based on Timestamping Authorities (TSA) and 
a Public Key Infrastructure (PKI), 
which addresses both object integrity and certification of an object's existence before a certain point in time. 
This specification
takes into account the phasing out of underlying algorithms by renewing integrity information
with newer timestamps obtained with newer keys or from a newer TSA. 
This solution however depends on the reliability of the TSAs used, which now become single points of failure 
for integrity information while they also remain outside the control of the digital preservation system.

Persistence was studied thoroughly by Driscoll et al in~\cite{STOC86*109}. Maniatis et al used the node copying technique
to introduce a persistent authenticated search tree in~\cite{SECURITY`02*297}, using a multi-key tree (B- Tree), which can be converted to a dictionary with
some additional work. However, space requirements quickly become prohibitive as any change in a snapshot requires complete B- Tree page copies.
Anagnostopoulos et al. introduced the Persistent Authenticated Dictionary (PAD) concept in~\cite{Anagnostopoulos:2001:PAD}, using a persistent Red-Black tree and node copying. More recently, in~\cite{crosby09thesis,conf/esorics/CrosbyW09} Crosby and Wallach studied the subject of PADs and suggested different techniques to implement them, including the treap~\cite{SeiAra96} with its set-unique property, as well a discussion about authenticator caching, which influenced our design decisions. However, this line of work never studied the impact of secondary memory on these data structures as all experiments were done in main memory.

In our work, we organize all integrity metadata in a single 
data structure, addressable by object name. We ensure this data structure is 
tamper-evident and we provide a complete solution to its preservation and 
recovery, as well as its own integrity checking. This way, the host digital
preservation system does not need to track any extra information alongside the objects. 
Additionally, when the host system asks for the integrity information of an object 
(by name), it is assured this information is intact. 
With our TreapPAD, our solution is space efficient, growing linearly with the number of elements in the dictionary,
while also giving solutions to managing the caching of authenticators.
Finally, our approach does not directly use a PKI; however it does need secure
message delivery to and from the nodes comprising the verifiers and preservers lists.
It is up to the host system to choose the right method for message authentication and integrity checks, 
however it does not have to store any secrets along the objects for the long term.

Our system is complementary to digital preservation and distributed storage systems and may be used 
as a tool by each storage node in the system to proactively verify the integrity of its contents.
As such, it is more suitable to systems where the complete file is stored in different nodes, such as
LOCKSS~{\cite{lockss-sosp-2003}}, FreeNet~{\cite{clarke00}}, 
FarSite~{\cite{Adya02}}, and Publius~{\cite{Waldman:2000:PRT}}. 
Systems that break files in pieces, such as Venti~{\cite{Quinlan:2002:VNA}},
OceanStore~{\cite{conf/asplos/KubiatowiczBCCEGGRWWWZ00}},
CFS~{\cite{conf/sosp/DabekKKMS01}},
Pastiche~{\cite{OSDI'02*285}},
Samsara~{\cite{SOSP'03*120}},
GridSharing~{\cite{conf/storagess/SubbiahB05}},
PASIS~{\cite{conf/dsn/GoodsonWGR04}},
and Glacier~{\cite{conf/nsdi/HaeberlenMD05}}
can still use IntegrityCatalog to proactively validate the pieces they own.

Our solution is orthogonal to external audits by the owner of the objects~{\cite{conf/usenix/LillibridgeEBBI03}}, or by an external 
auditor~{\cite{journals/iacr/ShahSB08}}. 

Finally, Muniswamy-Reddy et al. propose provenance information
should be tracked by creating metadata objects and treating them as first class objects~\cite{muniswamy2010provenance}; our system is complementary to theirs by taking over management of these metadata objects and assuring their full preservation. The same can be said for other provenance tracking systems, such as~\cite{hasan2009case}.

\makeatletter{}\section{Conclusion}
\label{sect:conclusion}

In this paper, we presented IntegrityCatalog, a novel distributed system that can be integrated into
digital repositories that use integrity tokens to periodically verify the
correctness of their locally stored data objects.  Unlike prior approaches
that focus only on making integrity metadata tamper-evident, IntegrityCatalog takes a
holistic approach.  It treats integrity metadata
as \emph{first-class} objects and provides a distributed network protocol
that ensures that integrity tokens are preserved, verified regularly, and recovered in case
of damage or adversarial attack, without requiring storage of long-term secrets.  Our approach uses a new data structure,
called TreapPAD, which provides a persistent, authenticated dictionary of
arbitrary length keys and associated values, that is space optimal and efficient
to synchronize over the network.  Measurements of our 
prototype after a full year's worth of operations, show 1K insertions per second, 
and 2K verified searches per second, as worst case performance.

Concluding, we believe IntegrityCatalog will prove useful to the digital preservation community. As future work, we plan to extend it to manage more provenance related information and integrate it with a provenance tracking system.

{\footnotesize \bibliographystyle{acm}
\bibliography{MyBibFile}}

\end{document}